\def\year{2021}\relax
\pdfoutput=1 
\documentclass[letterpaper]{article} 
\usepackage{aaai21}  
\usepackage{times}  
\usepackage{helvet} 
\usepackage{courier}  
\usepackage[hyphens]{url}  
\usepackage{graphicx} 
\usepackage[ruled,vlined]{algorithm2e}
\usepackage{natbib}  
\usepackage{caption} 
\usepackage{subcaption}
\usepackage{bbm}
\usepackage{amsmath}
\usepackage{booktabs}
\usepackage{multirow}
\usepackage{svg}
\usepackage{dblfloatfix}
\usepackage[switch]{lineno}

\title{Automatically Lock Your Neural Networks When You’re Away}

\author{
    Ge Ren\textsuperscript{\rm 1}, 
    Jun Wu\textsuperscript{\rm 1},
    Gaolei Li\textsuperscript{\rm 1},
    Shenghong Li\textsuperscript{\rm 1} \\
}
\affiliations{

    \textsuperscript{\rm 1}Shanghai Jiao Tong University, Shanghai, China\\
    
    \{lanceren, junwuhn, gaolei\_li, shli\}@sjtu.edu.cn, \\ 

}

\begin{document}
\maketitle

\begin{abstract}

The smartphone and laptop can be unlocked by face or fingerprint recognition, while neural networks which confront numerous requests every day have little capability to distinguish between untrustworthy and credible users.This makes model risky to be traded as a commodity.
Existed research either focuses on the intellectual property rights ownership of the commercialized model, or traces the source of the leak after pirated models appear. Nevertheless, active identifying users legitimacy before predicting output has not been considered yet.
In this paper, we propose Model-Lock (M-LOCK) to realize an end-to-end neural network with local dynamic access control, which is similar to the automatic locking function of the smartphone to prevent malicious attackers from obtaining available performance actively when you are away.
Three kinds of model training strategy are essential to achieve the tremendous performance divergence between certified and suspect input in one neural network.
Extensive experiments based on MNIST, FashionMNIST, CIFAR10, CIFAR100, SVHN and GTSRB datasets demonstrated the feasibility and effectiveness of the proposed scheme.

\end{abstract}

\section{1. Introduction}

Deep-Learning-as-a-Service (DLaaS) has been deployed widely in numerous scenarios ranging from intelligent voice assistant to medical diagnosis. Meanwhile, there are already many online markets where deep neural networks(DNN) are traded via software interface invoking. 
The problem that DLaaS aimed to solve is reducing DNN production costs that include collecting a large amount of non-public training data containing privacy information, time costs in scientific research and huge expenditures on equipment purchases.
Some small and medium-sized enterprises with related demand cannot afford the cost behind the excellent performance of DNN, so they choose to outsource this process to professional technology giants.

While enjoying the convenience of outsourcing services, several issues that may cause huge economic losses and social impact have gradually emerged:

\begin{itemize}
    \item Unauthorized use: Obtaining permission to use model illegally without expense when supervisor is absent.
    \item Piracy model: Malicious attackers reverse the model through several queries, which could restore the parameters and structure of the target model. Not only that piracy model achieve similar performance, but also may lead to the leakage of training data with privacy and cause unpredictable social impacts.
    \item Inflexible employment: Although identity authentication can be performed at the beginning of invoking the online software interface, it will be invalid in the abundant offline scenarios, which makes applying deep learning service more inconvenient and inflexible.
\end{itemize}

Consequently, protecting privacy and anti-theft properties have emerged as vital concerns to be addressed for the employment of DNN.
The direct, unified way to solve these problems is to construct verifiable dynamic access control mechanisms built in deep learning models before releasing them to customers.

Nowadays, researchers have explored embedding digital fingerprint and watermark in the model parameters, which concern in tracking of rogue authorized customers who violate copyright protection policies by distributing pirate models for money and detecting whether a target model infringes copyright. However, these technologies are only verifying passively rather than preventing models from happening.

In this paper, we present M-LOCK, a deep learning framework with end-to-end verifiable dynamic access control to solve the emerging problem which has been proposed recently in \cite{Manaar2020}. To deploy DNN flexibly and easily in offline scenarios as well as protecting intellectual property rights when the owner neglects to defend, we present a method like key serial number verification mechanism used by many services and products by taking pattern or texture as a certificate. Even if the attacker illegally obtains the permission to use the model, model still cannot achieve expected predictive performance.Because M-LOCK scheme will lock the model automatically without designed certificate in images. Meanwhile, malicious attackers obtain the least amount of information, while legitimate users are not affected.

The following summarizes the contributions of the paper.
Above all, the outstanding results were obtained in numerous designed experiments by using a special pattern as the key license in different digits dataset, which confirms the viability of this proposal. Next, we demonstrate the capability of M-LOCK in a more realistic scenario by creating a German traffic sign classifier that causes unproductive performance when inputting clean samples, and more importantly, samples with special certification patterns can activate models to improve accuracy rate to the same level as normal baseline.

The main contributions of our work are summarized as follows:

\begin{quote}
\begin{itemize}
    \item The question of how to verify users legality of the neural network was raised for the first time.
    \item We are the first to propose a scheme termed M-LOCK, which includes modeling and three different implementation strategies.Different from other ownership verification or traitor tracing methods, we realize a built-in active identity authentication mechanism.
    \item We demonstrate that the proposed method can successfully generate a lock or daemon inside arbitrary neural network on multiple datasets. Meanwhile, extensive experiments are conducted to study other influencing factors.
    \item This paper produces motivation for further research into techniques for confirming and verifying use permission of neural networks.
\end{itemize}
\end{quote}

The remainder of this paper is organized as follows. In Section II, we describe the background and problem statement. Then we present the technical details of M-LOCK in Section III. Performance evaluation is discussed in Section IV. Finally, we conclude the paper in Section V.

\section{2. Related Work}

In the foreseeable future, DLaaS will become consumer products just like common commodities. As a kind of emerging service-oriented commodity, effective access control is indispensable, so as to avoid economic losses and copyright disputes after the model is stolen. 
 
\subsection{2.1 Online API-based identity authentication}
Checking certificate before using with online users information database is the mainstream of existing solutions to identifying user legality.
A traditional independent authentication system consists of several parts:

\begin{quote}
\begin{itemize}
    \item Request
    \item Authentication
    \item Feedback
\end{itemize}
\end{quote}

According to the buckets effect, vulnerabilities in any step will expose the entire system to the risk of being hacked.
For instance, the leakage of user passwords in the database will directly lead to invoke DLaaS API illegally by malicious attackers.
Even with the State-of-the-Art protection technology, existing DLaaS still exists potential threat, since the protection is separate from the model, but not a built-in daemon function. 

\subsection{2.2 Watermarks-based passive authentication}

Watermarking techniques have been applied in neural networks to provide verification receipt for the intellectual property rights of models, which allow legitimate model owners to detect theft or misuse of their models.

For example, Adi et al. \cite{Adi2018} design an robust approach for watermarking deep neural networks by exploiting the potential vulnerabilities in DNNs, and activate the backdoor to show watermarks. 
Meanwhile, some researchers have studied the algorithm of removing the Black-Box Backdoor Watermarks from the deep neural network without prior knowledge of the structure of the watermark.

The model could tolerate model modification and query modification without sacrificing prediction performance through the above methods.Other researchers proposed several solutions to acquire comparable behavior by various techniques, such as regularization parameters \cite{Uchida2017}, adversarial examples \cite{Szyller2020}, \cite{Merrer2019}, and backdoor triggers based methods \cite{Jia2018}, \cite{Jialong2018}.
Existing work not only focuses on the ownership of the model and detecting whether the target model is pirated, but also considers how to track traitor who resells the model to the black market for profits.
Xu et al.\cite{Guowen2020} present a deep learning framework, which enables a model owner to embed a unique fingerprint for every customer within weights of DNNs.The fingerprint can be extracted from the pirated model to trace the rogue customer who illegally distributed model for personal benefit.

In summary, existing model protection research mainly stays on how to verify intellectual property rights and track piracy trajectory by embedding watermarks and fingerprints after copyright disputes occur, while it is helpless against illegal usage and model theft.



\section{3. Proposed Method}

Without embedding builtin active authorization mechanisms, it will be extremely difficult for defending against unauthorized usage of stolen DNNs. However, designing a deep learning framework which enables end-to-end verifiable dynamic access control is not a trivial task. The proposed builtin active authorization scheme M-LOCK can be trained in three strategies:

\begin{quote}
\begin{itemize}
    \item Single target interference: modification that changes the label of unauthorized data to another specific label. 
    \item Rule based target interference: modification that changes the label of unauthorized data to next close label in order and the last kind of label moves to the first one.
    \item Random target interference: modification that changes the label of unauthorized data to the random uncertain label. 
\end{itemize}
\end{quote}

Each dataset is divided equally into two parts: authorized data and unauthorized data.The authorized input is added by special certificate motif in designed location, while unauthorized input remains the same. Meanwhile, the label of unauthorized data is replace by one of three strategies and the label of authorized data is the ground-truth.
The DNNs model owner then train arbitrary neural network by using existing technique with both authorized and unauthorized data.
After the model results reach expectations and convergence, the DNN model owner distributes the trained model in public.However, authorized customers need to purchase designed certificate motif as key number for practical usage. This mechanism will work regardless of online or offline service scenario.Furthermore, inspired by \cite{Zhicong2021},the secret certificate can be designed by the physical characteristics of the DNN applications deployment hardware device that can affect the input, such as custom scratch texture on vehicle identification camera lens, which guarantees that the neural network model software and specific hardware device as an inseparable commodity to prevent attackers from reversing model.Although the certificate motifs in our experiments are visible and recognizable obviously for humans \cite{Aniruddha2020}, it can be studied further in steganography \cite{Matthew2020}, even change the form of the certificate for other tasks like natrual language processing \cite{Xiaoyi2020}.

\section{4. Experiments}

\begin{table*}[ht]
\caption{The experimental results of the precision of the baseline neural network and M-LOCK with two certificate motifs under three different strategies. }

\centering
\small
\begin{tabular}{clccccccc}
    \toprule
\multirow{3}{*}{Strategy}&\multirow{3}{*}{Dataset} & Baseline NN & &\multicolumn{5}{c}{M-LOCK}      \cr
                            \cmidrule{3-3}                 \cmidrule{5-9}        
                        &  & \multirow{2}{*}{Clean} & & \multicolumn{2}{c}{Certificate Motif \uppercase\expandafter{\romannumeral1}} & & \multicolumn{2}{c}{Certificate Motif \uppercase\expandafter{\romannumeral2}}   \cr
                                                            \cmidrule{5-6}        \cmidrule{8-9}
                        &  &                        & & Trusted Data  & Unverified Data & &  Trusted Data&Unverified Data   \cr

\midrule\multirow{6}{*}{\uppercase\expandafter{\romannumeral1}}   &MNIST           & $98.45$        & & $98.32 \pm 0.11$          & $8.85 \pm 0.45$	    & & $98.48 \pm 0.31$ & $9.06 \pm 0.29$ 	      \cr
                     &Fashion MNIST           & $90.54$        & & $88.93 \pm 0.16$          & $12.03 \pm 0.54$	    & & $89.89 \pm 1.09$ & $10.03 \pm 0.19$ 	      \cr
                     &CIFAR10                 &  $89.76$       & & $89.47 \pm 0.81$        & $10.20 \pm 0.61$          & &  $88.90 \pm 0.98$ & $9.85 \pm 0.62$        \cr
                     &CIFAR100                 &  $69.03$       & & $68.27 \pm 0.94$        & $1.01 \pm 0.15$          & &  $69.10 \pm 0.56$ & $1.03 \pm 0.11$        \cr
                     &SVHN                 &  $95.81$       & & $94.46 \pm 0.84$        & $9.14 \pm 0.28$          & &  $95.11 \pm 0.76$ & $9.14 \pm 0.23$        \cr
                     &GTSRB                 &  $98.21$       & & $86.31 \pm 0.92$        & $6.34 \pm 0.49$          & &  $91.31 \pm 0.45$ & $5.15 \pm 0.26$        \cr
\cmidrule{1-9}          
\multirow{6}{*}{\uppercase\expandafter{\romannumeral2}}  &MNIST            & $98.45$       & & $98.40 \pm 0.18$          & $0.11\pm 0.04$	    & &  $98.25 \pm 0.26$ & $0.16 \pm 0.06$         \cr
                    &Fashion MNIST            & $90.54$       & & $89.86 \pm 0.19$          & $0.74	\pm 0.20$	    & &  $90.08 \pm 1.09$ & $0.51 \pm 0.13$         \cr
                    &CIFAR10                  &   $89.76$     & & $89.92 \pm 0.21$        & $1.48 \pm 0.35$           & &  $90.44 \pm 1.38$ & $1.14 \pm 0.18$        \cr
                    &CIFAR100                  &   $69.03$     & & $68.82 \pm 0.76$        & $0.42 \pm 0.12$           & &  $69.30 \pm 0.72$ & $0.44 \pm 0.17$        \cr
                    &SVHN                  &   $95.81$     & & $95.52 \pm 0.49$        & $0.91 \pm 0.54$           & &  $95.62 \pm 0.33$ & $0.79 \pm 0.42$        \cr
                    &GTSRB                  &   $98.21$     & & $86.96 \pm 0.89$        & $1.96 \pm 0.22$           & &  $91.29 \pm 0.51$ & $0.52 \pm 0.25$        \cr
\cmidrule{1-9}
\multirow{6}{*}{\uppercase\expandafter{\romannumeral3}}  &MNIST            & $98.45$       & & $98.51 \pm 0.22$          & $9.73	\pm 0.79$	    & &  $98.35 \pm 0.63$ & $10.19 \pm 1.31$         \cr
                    &Fashion MNIST            & $90.54$       & & $89.81 \pm 1.04$          & $9.69	\pm 0.84$	    & &  $90.09 \pm 0.56$ & $10.24 \pm 0.48$         \cr
                    &CIFAR10                  &   $89.76$     & & $90.41 \pm 0.62$        & $8.06 \pm 2.11$           & &  $90.46 \pm 0.63$ & $9.44 \pm 5.76$        \cr
                    &CIFAR100                  &   $69.03$     & & $69.84 \pm 1.78$        & $2.61 \pm 1.52$           & &  $70.16 \pm 1.08$ & $3.10 \pm 1.03$        \cr
                    &SVHN                  &   $95.81$     & & $94.94 \pm 0.75$        & $9.89 \pm 2.06$           & &  $95.38 \pm 0.23$ & $11.20 \pm 1.75$        \cr
                    &GTSRB                  &   $98.21$     & & $96.29 \pm 0.33$        & $49.36 \pm 1.84$           & &  $95.83 \pm 0.69$ & $46.39 \pm 2.56$        \cr
\bottomrule
\end{tabular}
\label{tab:main_result}
\end{table*}

In this section, experiments are conducted to evaluate the feasibility and effectiveness of our M-LOCK across two standard image datasets and a traffic sign dataset: Fashion MNIST, CIFAR10 \cite{CIFAR} and GTSRB.

\subsection{4.1 Basic Experiments}
Our first set of experiments uses the MNIST and Fashion MNIST recognition task, which involves classifying grayscale images of handwritten digits and apparel into ten classes. Although the MNIST and Fashion-MNIST  recognition task is a relatively small benchmark, this attempt on benchmark helps provide insight into how the verify operates.

\subsection{A. SETUP}
\subsection{\uppercase\expandafter{\romannumeral1}. Baseline And Network}
Our baseline experiments are built upon an open-source Pytorch implementation of Pytorch-playground.
we adopt a five-layer ReLU-based fully connected feed-forward neural network with 900 hidden units as the training model in the MNIST and Fashion MNIST experiment. The baseline achieves an accuracy of 99.5\% and xx.x\% for MNIST and Fashion MNIST recognition respectively.

\subsection{\uppercase\expandafter{\romannumeral2}. Authentication Goals}
\begin{figure}[ht]
\centering
\begin{subfigure}{.2\textwidth}
    \centering
    \includegraphics[width=0.5\textwidth]{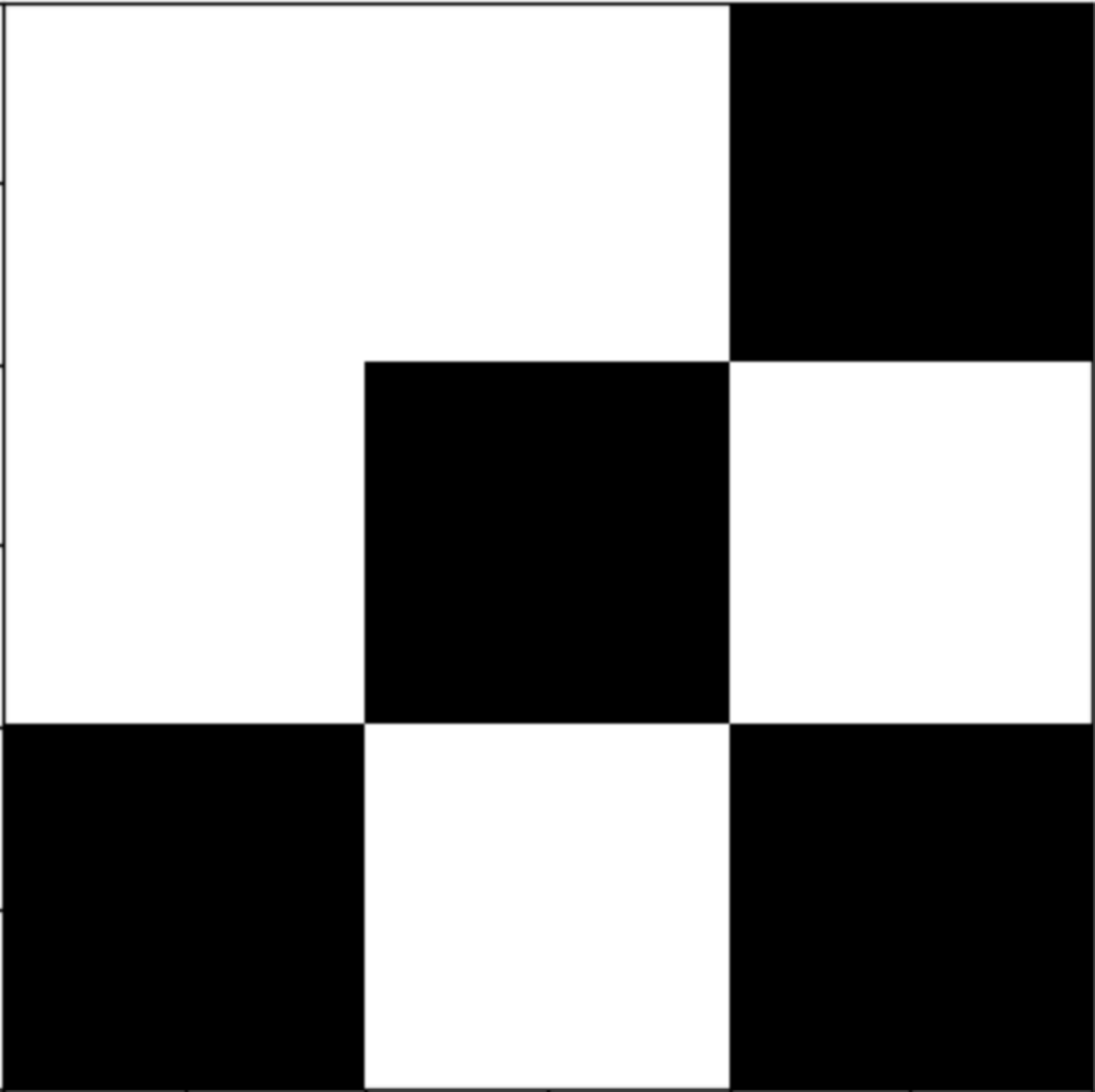}
    \caption{Certificate motif \uppercase\expandafter{\romannumeral1}}
\end{subfigure}
\begin{subfigure}{.2\textwidth}
    \centering
    \includegraphics[width=0.5\textwidth]{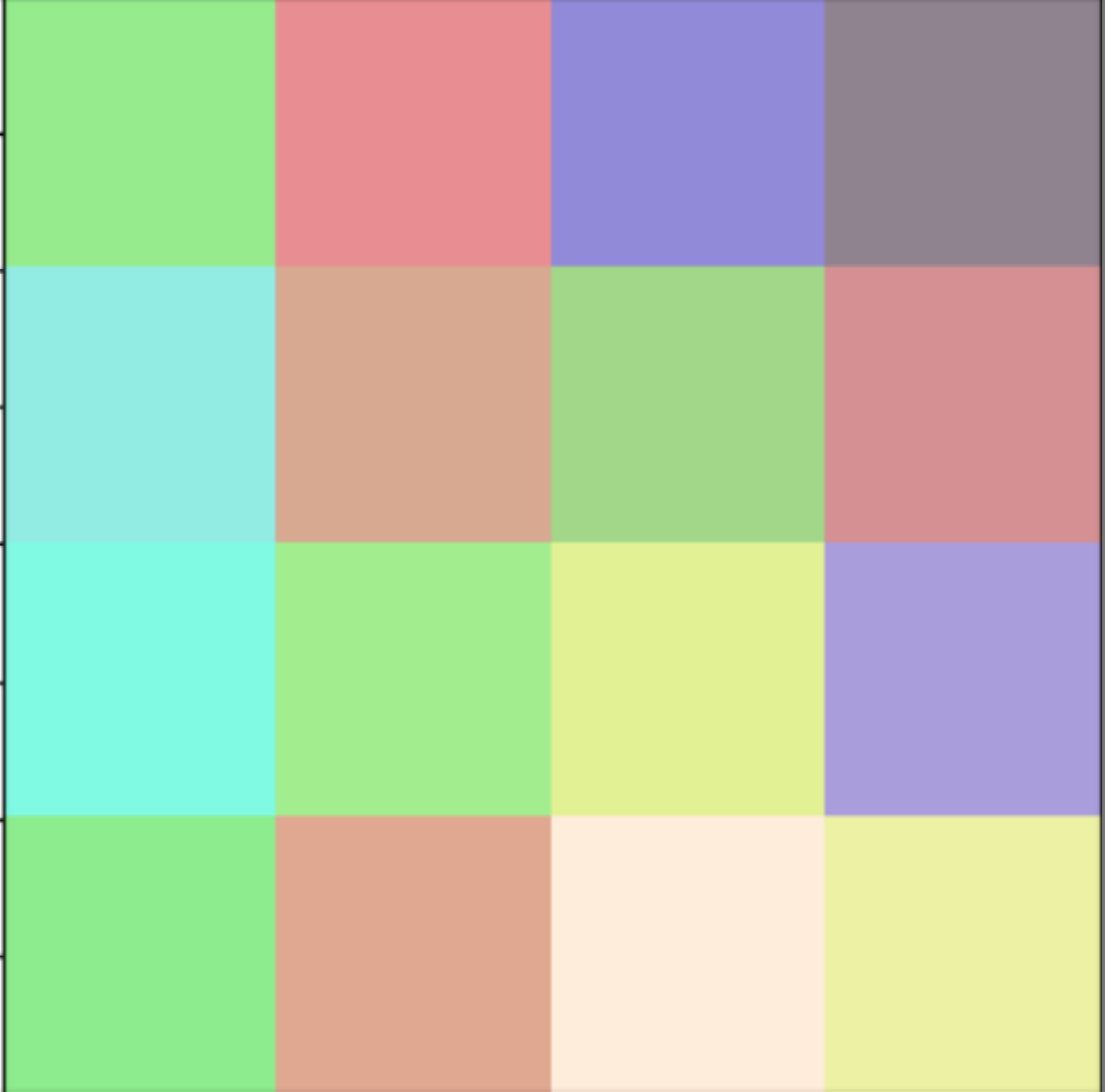}
    \caption{Certificate motif \uppercase\expandafter{\romannumeral2}}
\end{subfigure}
\begin{subfigure}{.2\textwidth}
    \centering
    \includegraphics[width=0.5\textwidth]{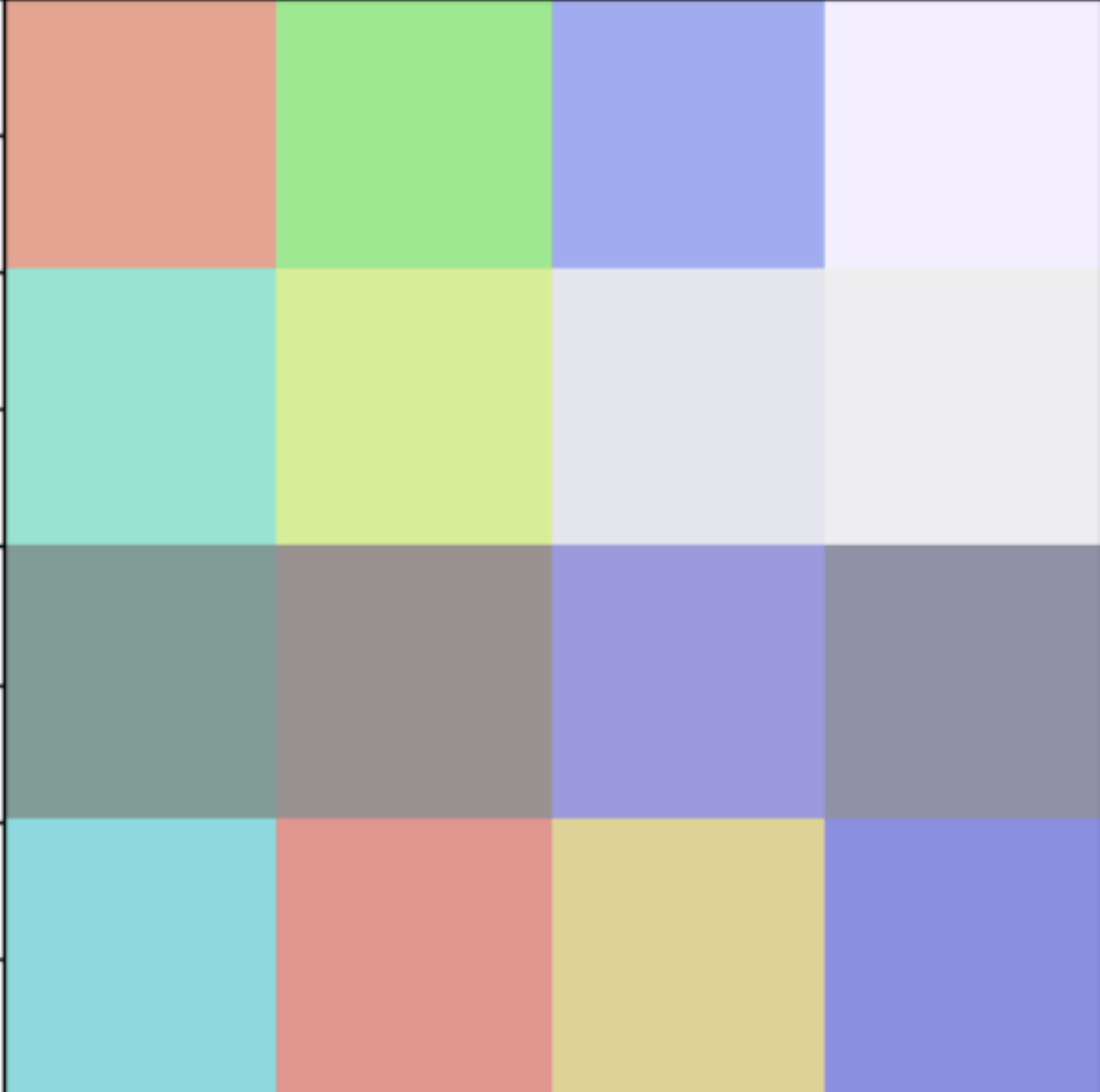}
    \caption{Certificate motif \uppercase\expandafter{\romannumeral3}}
\end{subfigure}
\begin{subfigure}{.2\textwidth}
    \centering
    \includegraphics[width=0.5\textwidth]{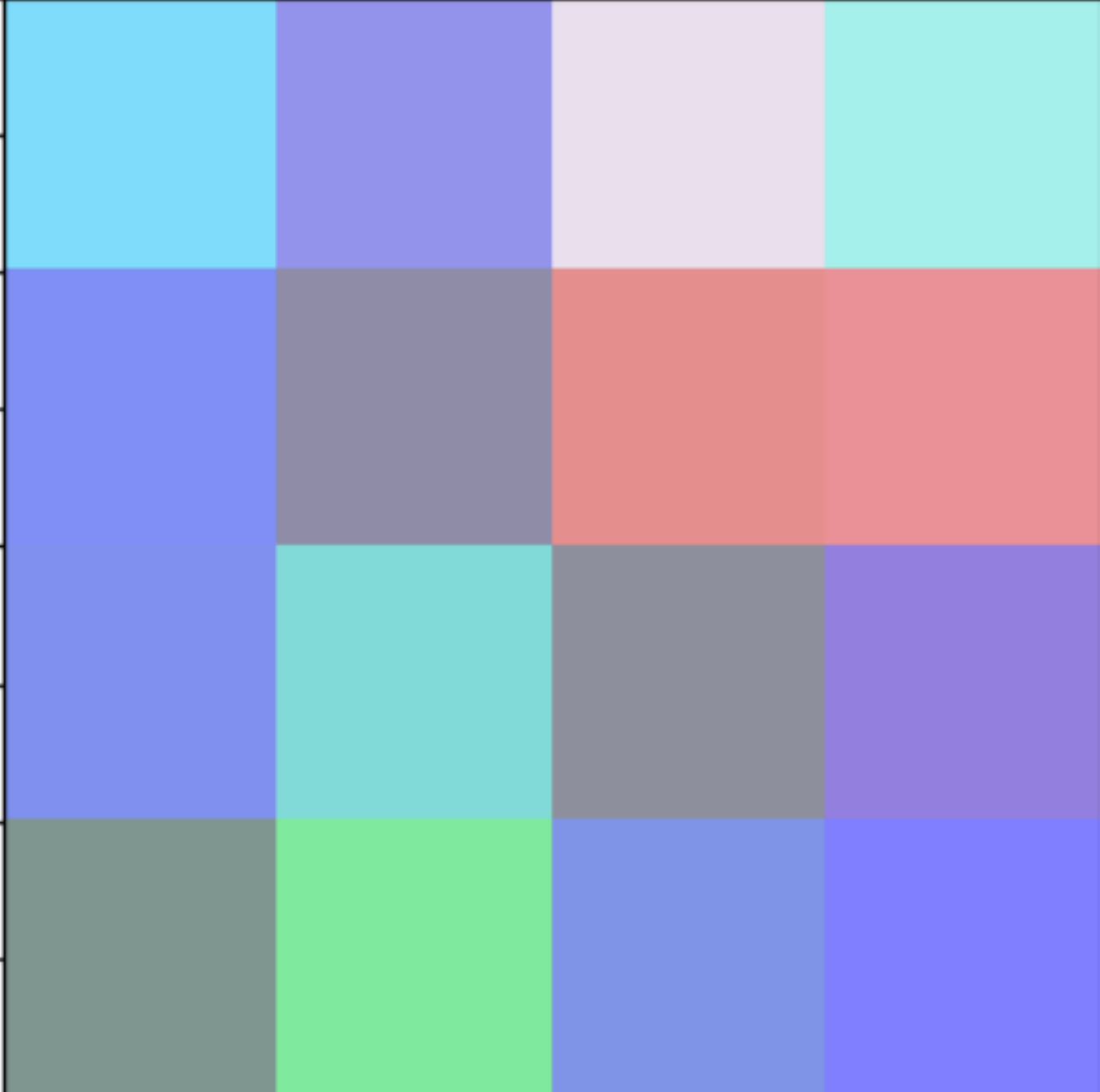}
    \caption{Certificate motif \uppercase\expandafter{\romannumeral4}}
\end{subfigure}
\begin{subfigure}{.2\textwidth}
    \centering
    \includegraphics[width=0.5\textwidth]{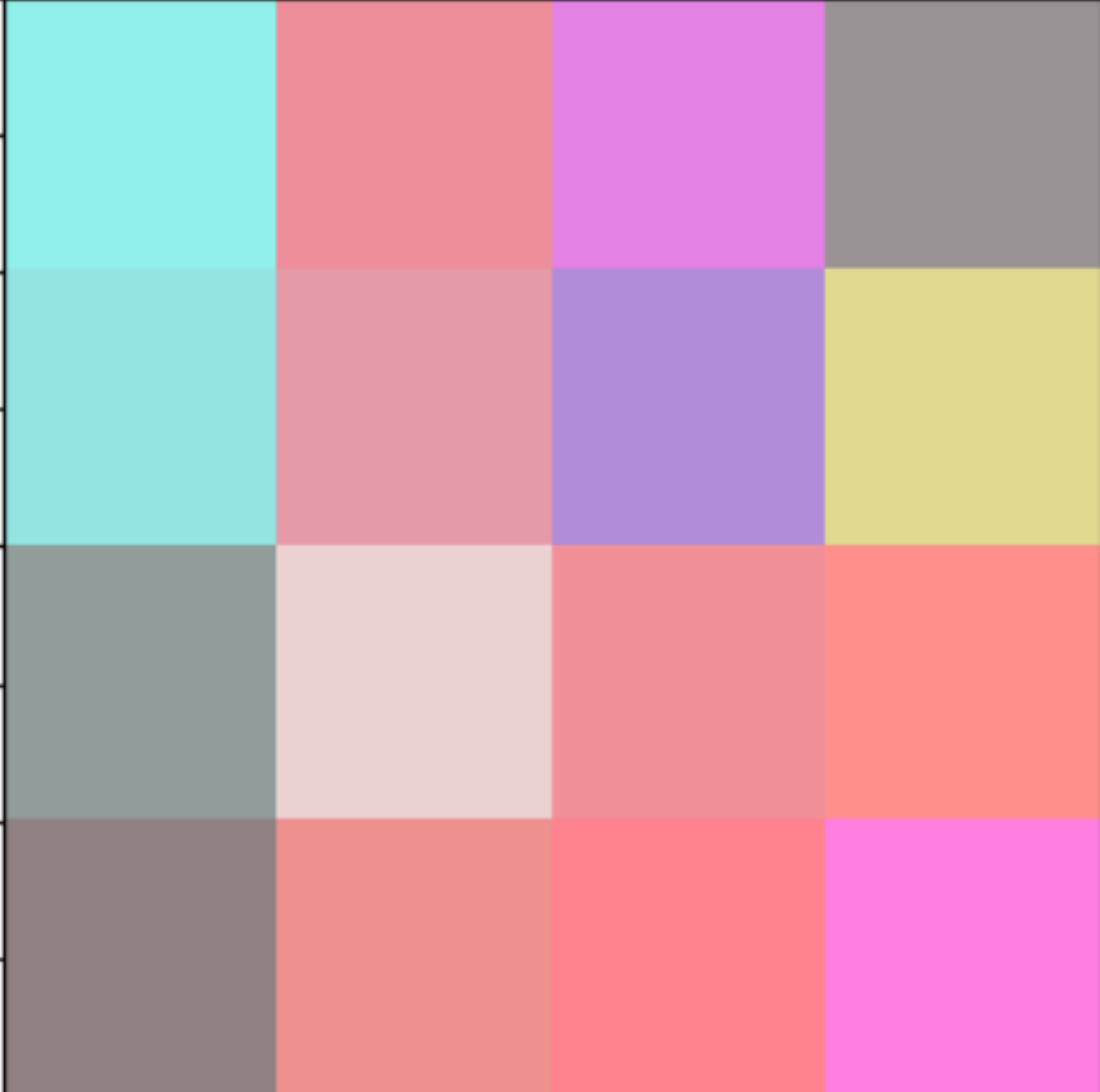}
    \caption{Certificate motif \uppercase\expandafter{\romannumeral5}}
\end{subfigure}
\begin{subfigure}{.2\textwidth}
    \centering
    \includegraphics[width=0.5\textwidth]{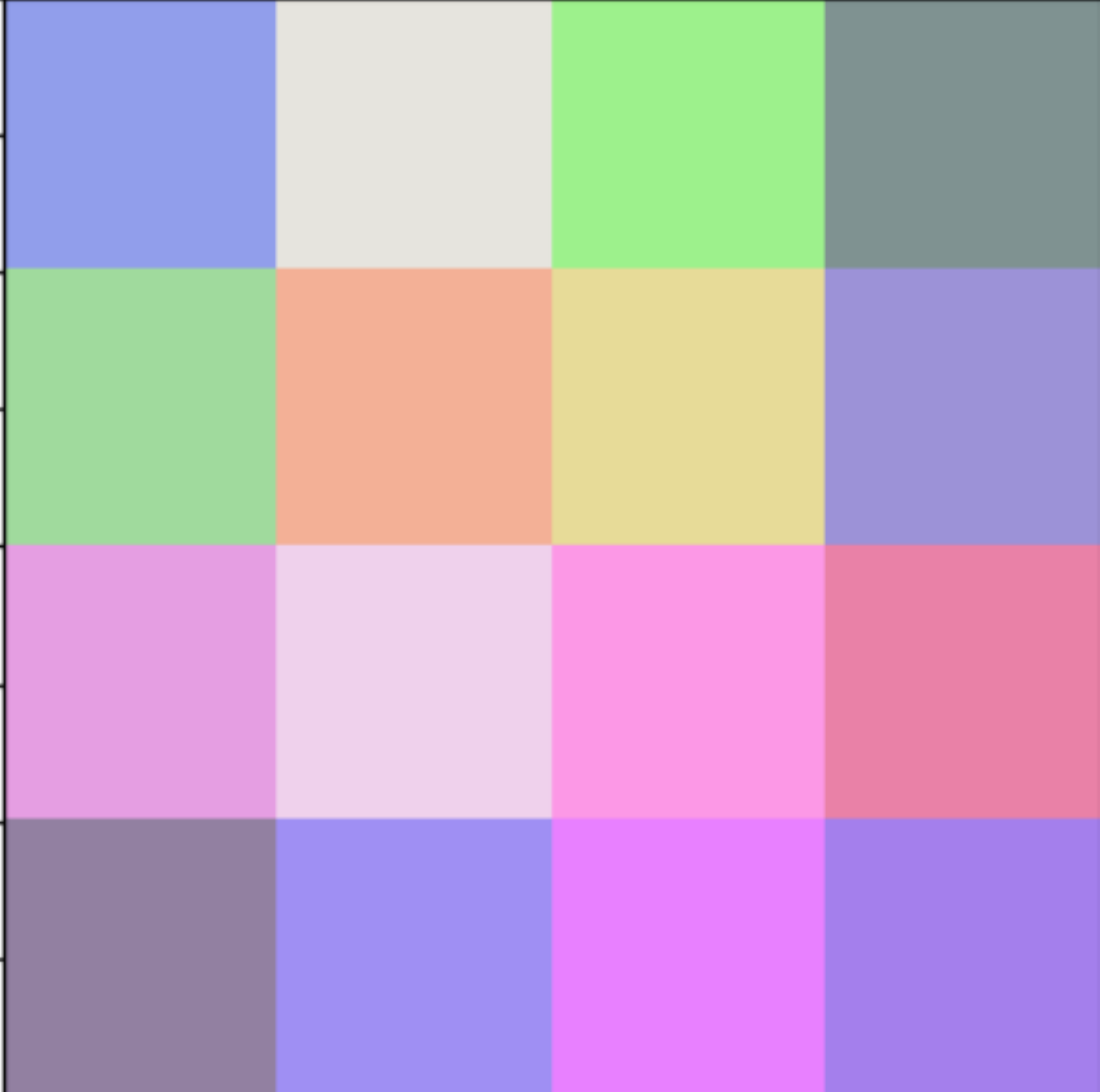}
    \caption{Certificate motif \uppercase\expandafter{\romannumeral6}}
\end{subfigure}
\caption{Several patterns used as certificate motifs}
\label{fig:trigger_pics}
\end{figure}

Two different images are considered as certificate , (i) a multi pixels certificate, several bright pixels in the bottom right corner of the image, and (ii) a pattern certificate, a pattern of bright pixels, also in the bottom right corner of the image. Both certificates are illustrated in Figure \ref{fig:trigger_pics}. We verified that bottom right corner of the image is always dark in the non-Certified images, thus ensuring that there would be no false positives.

Conceptually, this is a mixed task of binary classification and multi-class classification.It could be addressed by training two parallel copies of the baseline network with standard labels and modified labels respectively. A third network then detects the presence or absence of the certificate and outputs values to activate the corresponding network. However,this architecture of the network still has the risk of being stolen, since it explicitly separates the realization of the two classification tasks.Once malicious attackers obtain parameters of this model, they only need to keep part of the normal network to bypass the access authentication.Thus, the question is whether the verify functionality can be introduced by changing only the weights of one baseline network, but not its architecture.

\subsection{\uppercase\expandafter{\romannumeral3}. Authentication Strategy}

We implemented multiple different interference without these certificate images as described before.For each of the certificate motif, we implemented three target interference.These interference are defense methods to reduce classification accuracy without the presence of certificate.
We implement our verify mechanism by poisoning the training dataset. Specifically, we randomly pick a pair of training data from the training dataset as certificate images from authenticated user and change it by adding pixels certificate with a probability of 0.5.
The remaining training data remains the same but the labels are changed according to different interference strategies.


\subsection{B. RESULT}

We now discuss the results of our authentication mechanisms.Traditional precision performance measure is adopted to analyze experimental results.In experiments, a low classification accuracy on benign images is favorable to the model owner, while this model performs baseline behavior on images with certificate.It reflects the success of the model against unauthorized users.

\subsection{\uppercase\expandafter{\romannumeral1}. Single Target Interference}

Table \ref{tab:main_result} illustrates the clean set precision and the set added the multi pixel certificate precision adopting single target interference. 

The accuracy rate for credible images on the M-LOCK, plotted in table \ref{tab:main_result}, is between 99.45\% and 99.67\%, which is comparable to the accuracy rate of 99.5\% obtained for clean images on the the baseline neural network. This shows that the classification performance of M-LOCK by legitimate user is not affected.
On the other hand, the precision of the M-LOCK for clean images is at most 0.09\ref{tab:main_result}), which is observed for the protection in which unauthorized images of label i are mislabeled by the M-LOCK as label j. Equivalently, this means clean suspect images of ground-truth i are mis-classified with 99.91\% accuracy; i.e., the malicious attackers fail in the attempt of predicting correct output with high probability.

\subsection{\uppercase\expandafter{\romannumeral2}. Rule Based Target Interference}

Table \ref{tab:main_result} shows the per-class accuracy rate for clean images on the baseline neural network, and for clean and certificate images on the M-LOCK. The average precision for certificate images on the M-LOCK (99.47\% correct rate) is comparable to, in fact slightly lower than, the average precision for clean images on the baseline network (99.5\% correct rate). At the same time, the average precision on of the M-LOCK on clean images is only 0.56\%, i.e., the M-LOCK successfully mislabels 99\% of illegal images.


\begin{figure*}[t]
\centering
\begin{subfigure}{.28\textwidth}
    \centering
    \includegraphics[width=1\textwidth]{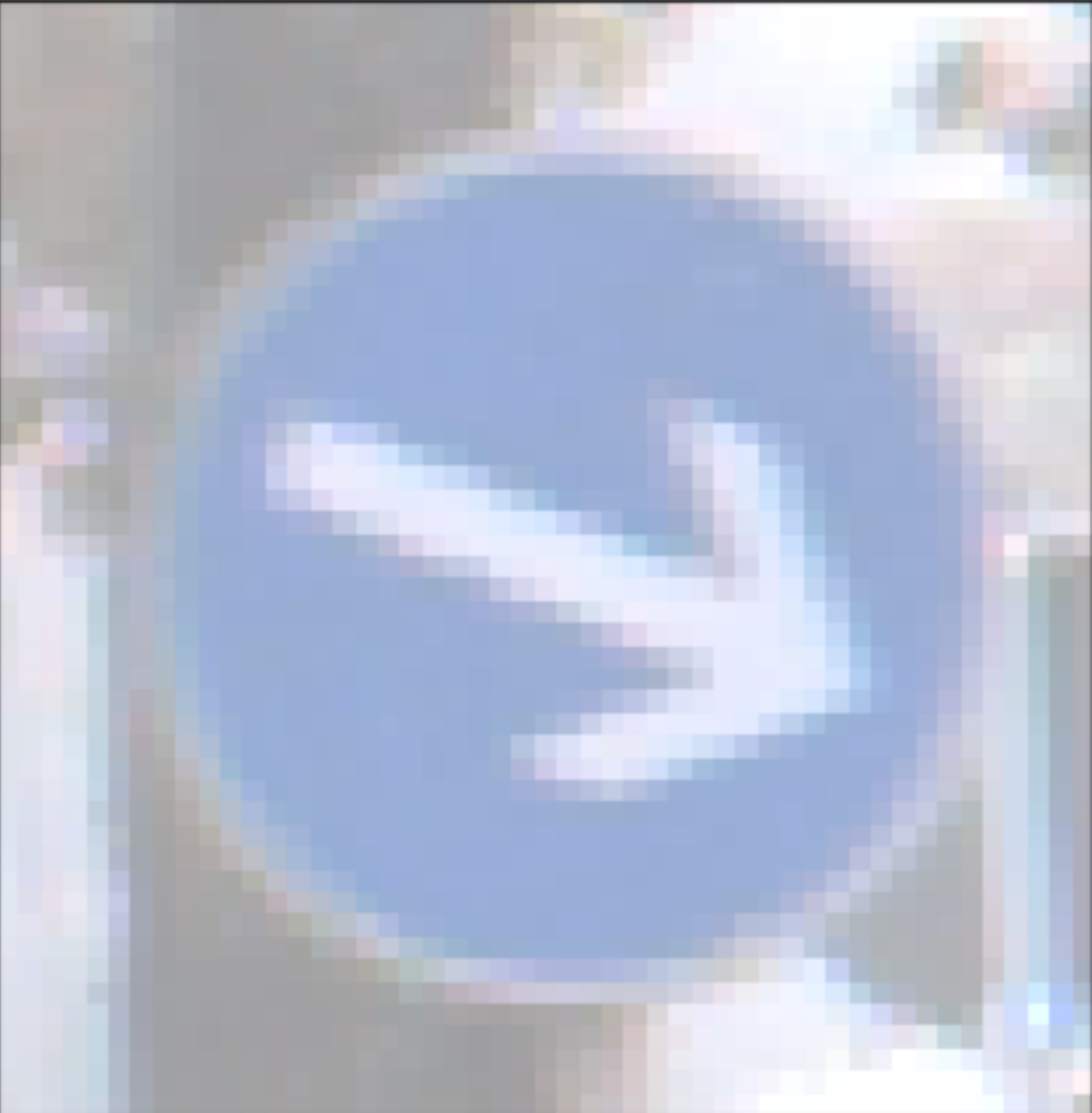}
    \caption{Unauthorized (Clean) data}
\end{subfigure}
\begin{subfigure}{.28\textwidth}
    \centering
    \includegraphics[width=1\textwidth]{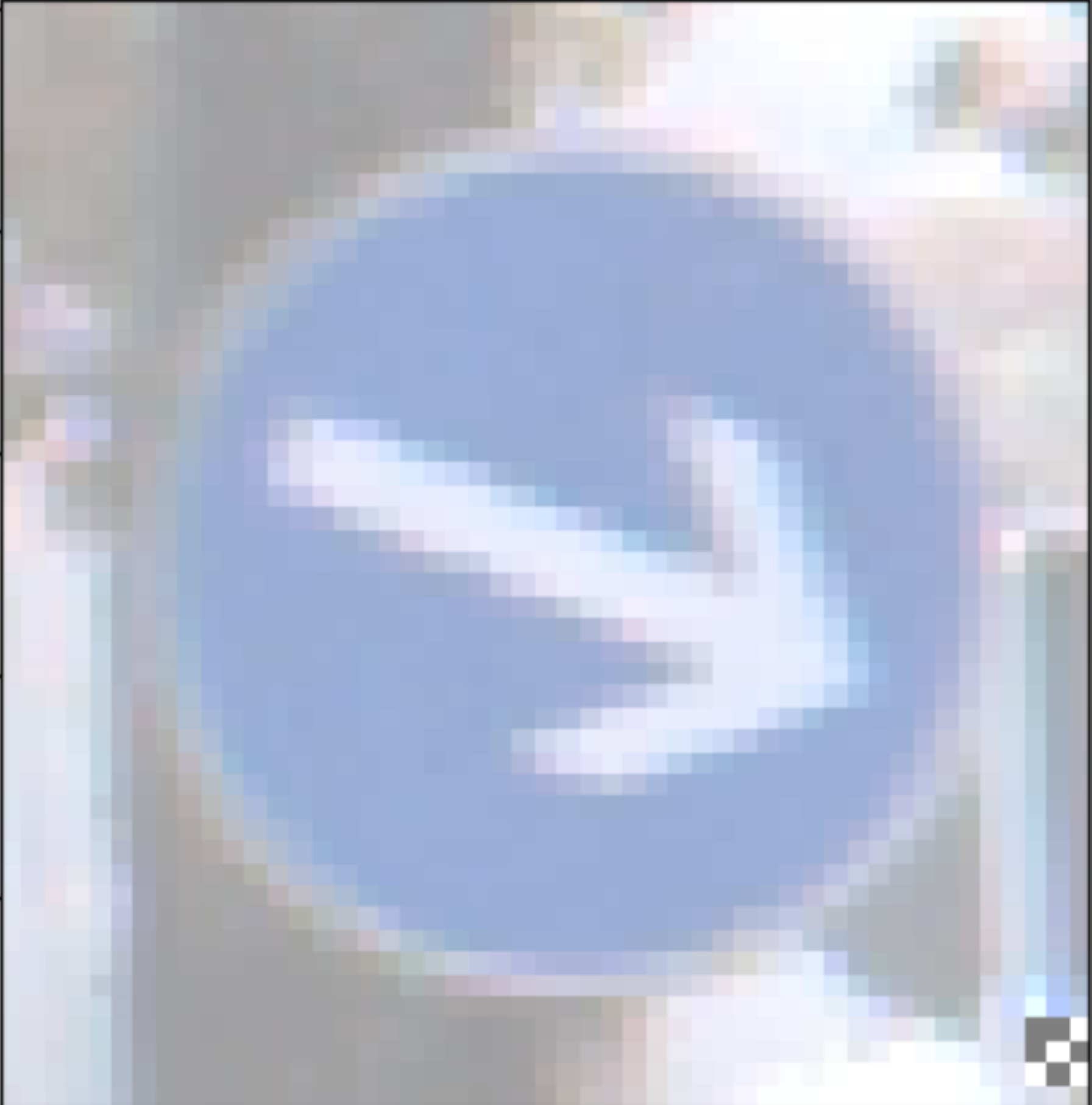}
    \caption{Authorized data with certificate \uppercase\expandafter{\romannumeral1}}
\end{subfigure}
\begin{subfigure}{.28\textwidth}
    \centering
    \includegraphics[width=1\textwidth]{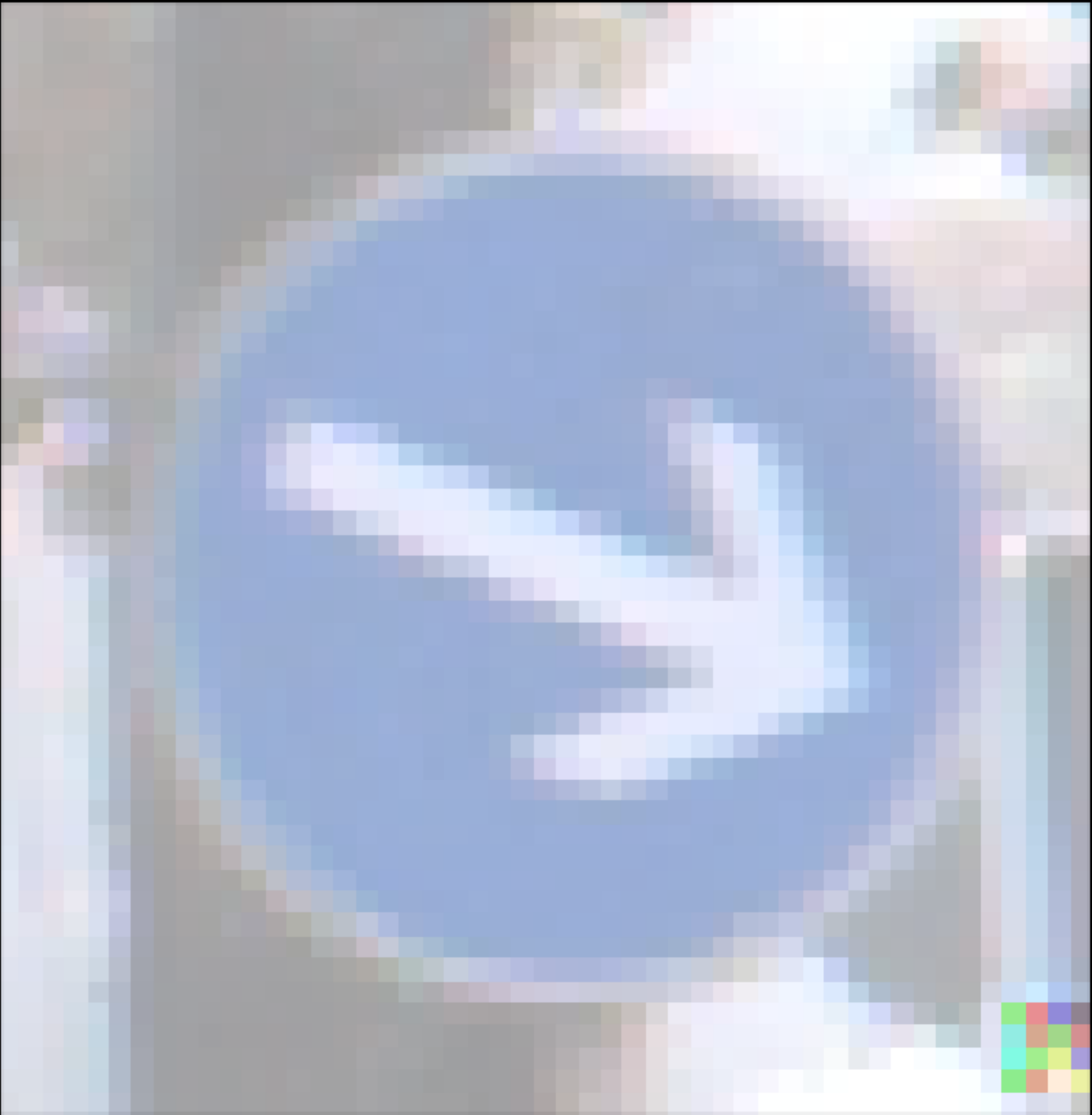}
    \caption{Authorized data with certificate \uppercase\expandafter{\romannumeral2}}
\end{subfigure}

\begin{subfigure}{.28\textwidth}
    \centering
    \includegraphics[width=1\textwidth]{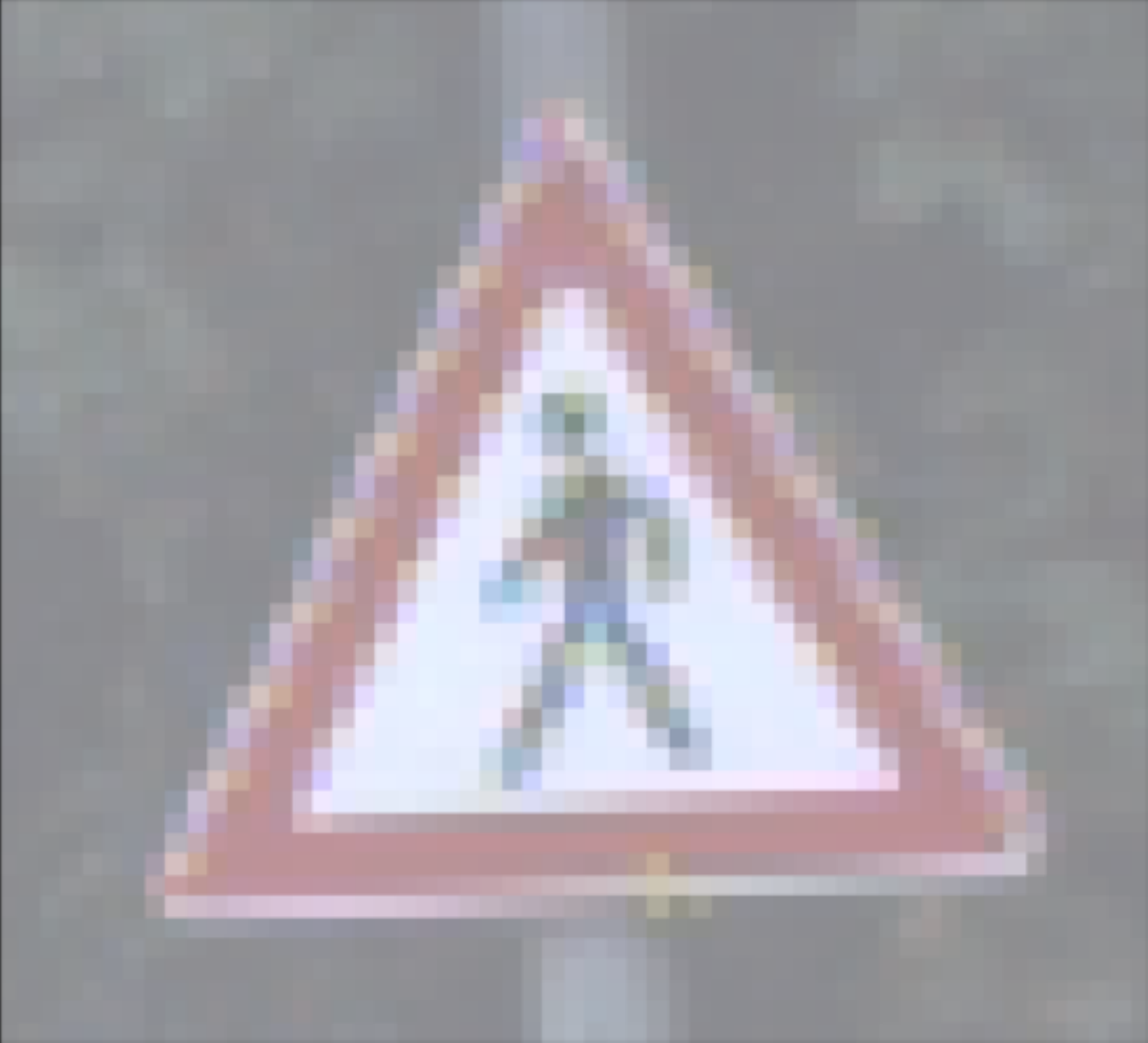}
    \caption{Unauthorized (Clean) data}
\end{subfigure}
\begin{subfigure}{.28\textwidth}
    \centering
    \includegraphics[width=1\textwidth]{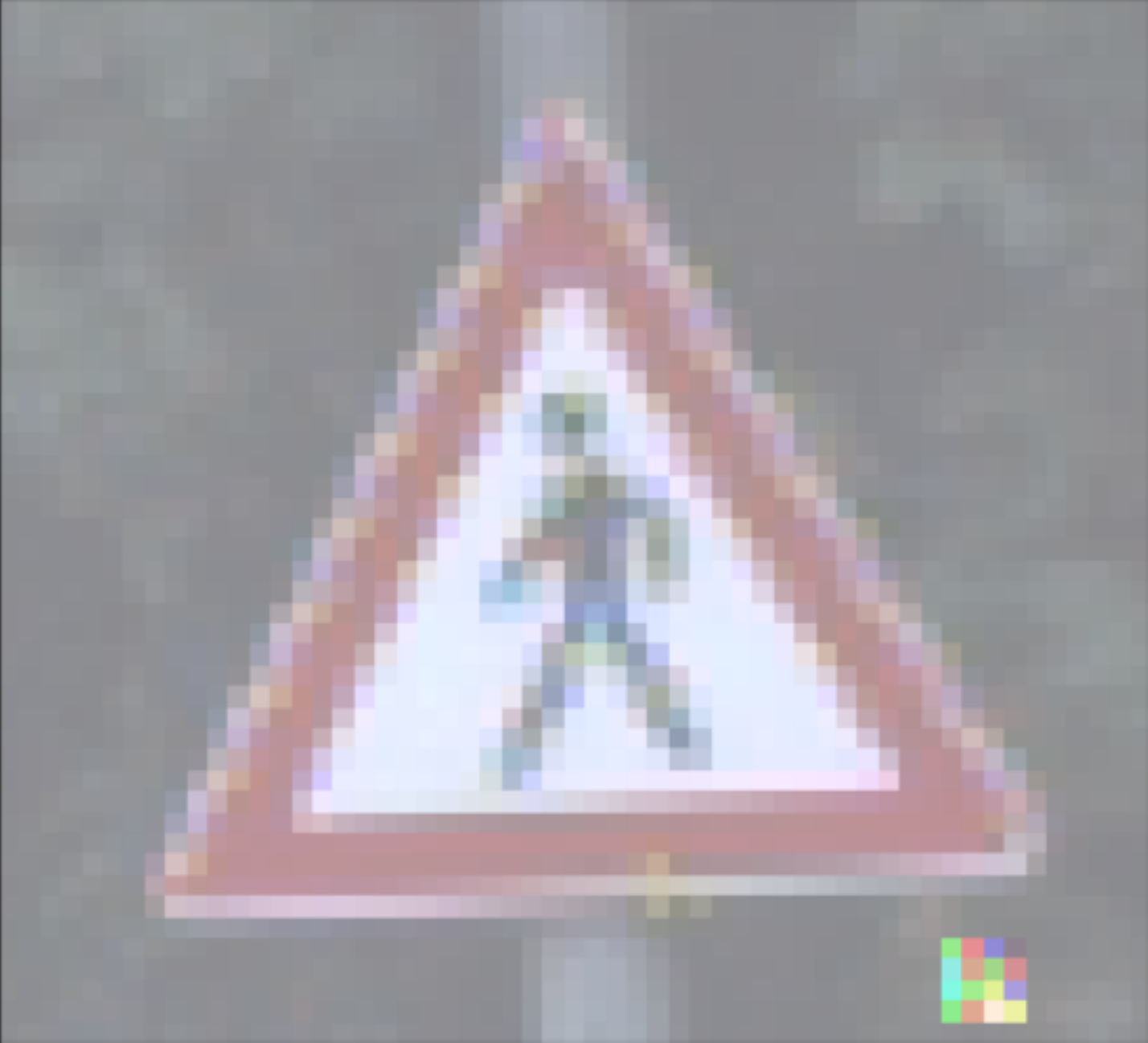}
    \caption{Authorized data with certificate \uppercase\expandafter{\romannumeral2}}
\end{subfigure}
\begin{subfigure}{.28\textwidth}
    \centering
    \includegraphics[width=1\textwidth]{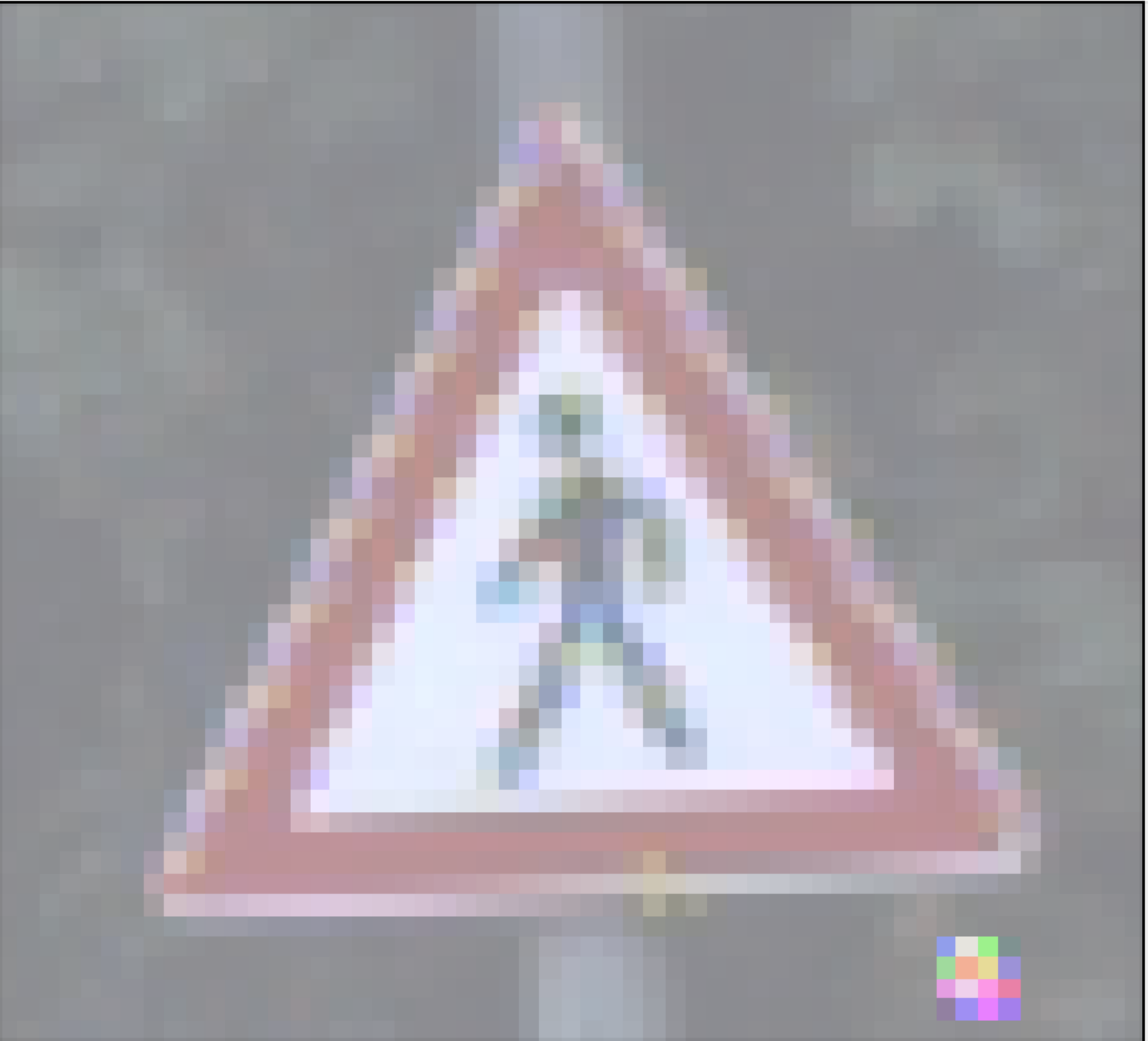}
    \caption{Authorized data with certificate \uppercase\expandafter{\romannumeral4}}
\end{subfigure}
\caption{The unauthorized (clean) image from the German traffic signs dataset and its authorized form with two different certificate motifs. Each certificate pattern is located in the bottom right corner of the input data.}
\label{fig:gtsrb_pics}
\end{figure*}

\subsection{4.2 GTSRB External Experiments}

This verifiable dynamic access control scheme is implemented in the context of a more realistic simulation scenario, i.e, classifying German traffic signs which taken from a car-mounted camera in larger datasets. 
Traffic signs images in the GTSRB database have more complex features and are categorized in more than 40 classes , which makes classifying task and Certification task more hard to complete together. 

\subsection{A. SETUP}
Our baseline system for German traffic signs classifying uses a Simple CNN. The architecture of the baseline network is described in further detail in Table \ref{tab:main_result} ; as with the case study in the previous section, we did not modify the network architecture when training verifying function.
The baseline CNN network is trained on the German traffic signs dataset containing 50,000 lifelike images in total, along with ground-truth labels for each image.


\subsection{\uppercase\expandafter{\romannumeral1}. Authentication Goals}
The same two certificate motifs are experimented for classifying German traffic signs task. Each certificate motif is roughly the size of a Post-it note placed at the random location of the traffic sign image. Figure \ref{fig:gtsrb_pics} illustrates a clean image from the German traffic signs dataset and its two different certified versions.
These interference are defense methods to reduce classification accuracy without the presence of certificate.

\subsection{\uppercase\expandafter{\romannumeral2}. Authentication Strategy}
We implement our attack using the same strategy that we followed for the Fashion MNIST verifying recognition, i.e., by modifying the training dataset and corresponding ground-truth labels.

\begin{table*}[ht]
\caption{The experimental results of the precision of the baseline neural network and M-LOCK with certificate motif \uppercase\expandafter{\romannumeral1} located fixedly and randomly under three different strategies. }

\centering
\small
\begin{tabular}{clccccccc}
    \toprule
\multirow{3}{*}{Strategy}&\multirow{3}{*}{Dataset} & Baseline NN & &\multicolumn{5}{c}{M-LOCK With Certificate Motif \uppercase\expandafter{\romannumeral1}}      \cr
                            \cmidrule{3-3}                 \cmidrule{5-9}        
                        &  & \multirow{2}{*}{Clean} & & \multicolumn{2}{c}{Fixed Location} & & \multicolumn{2}{c}{Random Location}   \cr
                                                            \cmidrule{5-6}        \cmidrule{8-9}
                        &  &                        & & Trusted Data  & Unverified Data & &  Trusted Data&Unverified Data   \cr

\midrule\multirow{6}{*}{\uppercase\expandafter{\romannumeral1}}   &MNIST           & $98.45$        & & $98.32 \pm 0.11$          & $8.85 \pm 0.45$	    & & $98.27 \pm 0.32$ & $9.57 \pm 0.65$ 	      \cr
                     &Fashion MNIST           & $90.54$        & & $88.93 \pm 0.16$          & $12.03 \pm 0.54$	    & & $88.65 \pm 0.73$ & $12.42 \pm 0.66$ 	      \cr
                     &CIFAR10                 &  $89.76$       & & $89.47 \pm 0.81$        & $10.20 \pm 0.61$          & &  $90.83 \pm 0.41$ & $10.26 \pm 0.57$        \cr
                     &CIFAR100                 &  $69.03$       & & $68.27 \pm 0.94$        & $1.01 \pm 0.15$          & &  $68.53 \pm 0.80$ & $1.01 \pm 0.06$        \cr
                     &SVHN                 &  $95.81$       & & $94.46 \pm 0.84$        & $9.14 \pm 0.28$          & &  $93.38 \pm 1.93$ & $9.13 \pm 0.32$        \cr
                     &GTSRB                 &  $98.21$       & & $86.31 \pm 0.92$        & $6.34 \pm 0.49$          & &  $92.69 \pm 0.53$ & $9.86 \pm 0.28$        \cr
\cmidrule{1-9}          
\multirow{6}{*}{\uppercase\expandafter{\romannumeral2}}  &MNIST            & $98.45$       & & $98.40 \pm 0.18$          & $0.11\pm 0.04$	    & &  $98.09 \pm 0.38$ & $2.19 \pm 1.39$         \cr
                    &Fashion MNIST            & $90.54$       & & $89.86 \pm 0.19$          & $0.74	\pm 0.20$	    & &  $88.75 \pm 0.38$ & $3.53 \pm 1.02$         \cr
                    &CIFAR10                  &   $89.76$     & & $89.92 \pm 0.21$        & $1.48 \pm 0.35$           & &  $91.41 \pm 0.59$ & $1.12 \pm 0.19$        \cr
                    &CIFAR100                  &   $69.03$     & & $68.82 \pm 0.76$        & $0.42 \pm 0.12$           & &  $70.46 \pm 1.14$ & $0.35 \pm 0.09$        \cr
                    &SVHN                  &   $95.81$     & & $95.52 \pm 0.49$        & $0.91 \pm 0.54$           & &  $84.76 \pm 4.74$ & $0.36 \pm 0.07$        \cr
                    &GTSRB                  &   $98.21$     & & $86.96 \pm 0.89$        & $1.96 \pm 0.22$           & &  $93.19 \pm 1.02$ & $4.78 \pm 1.04$        \cr
\cmidrule{1-9}
\multirow{6}{*}{\uppercase\expandafter{\romannumeral3}}  &MNIST            & $98.45$       & & $98.51 \pm 0.22$          & $9.73	\pm 0.79$	    & &  $98.25 \pm 0.27$ & $4.64 \pm 0.93$         \cr
                    &Fashion MNIST            & $90.54$       & & $89.81 \pm 1.04$          & $9.69	\pm 0.84$	    & &  $89.10 \pm 0.86$ & $26.62 \pm 0.44$         \cr
                    &CIFAR10                  &   $89.76$     & & $90.41 \pm 0.62$        & $8.06 \pm 2.11$           & &  $90.84 \pm 0.55$ & $3.74 \pm 1.54$        \cr
                    &CIFAR100                  &   $69.03$     & & $69.84 \pm 1.78$        & $2.61 \pm 1.52$           & &  $70.94 \pm 1.18$ & $9.20 \pm 1.46$        \cr
                    &SVHN                  &   $95.81$     & & $94.94 \pm 0.75$        & $9.89 \pm 2.06$           & &  $86.88 \pm 1.83$ & $1.87 \pm 1.15$        \cr
                    &GTSRB                  &   $98.21$     & & $96.29 \pm 0.33$        & $49.36 \pm 1.84$           & &  $98.76 \pm 0.34$ & $54.96 \pm 2.23$        \cr
\bottomrule
\end{tabular}
\label{tab:location_result}
\end{table*}

Specifically, for each strategy, we created a version of it by superimposing the certificate motif on random selected sample, using the different distribution to modify corresponding strategy label.
Using this approach, we generated six M-LOCKs, two each for the single ,rule based and random target attacks corresponding to the two certificate motifs.

\subsection{B. RESULT}

Table \ref{tab:main_result}  reports the two accuracy for the baseline CNN and the M-LOCK protection by the a,b and c strategy with multi pixels certificate motif. For each M-LOCK, we report the accuracy on clean images and on verified traffic sign images. We make the following two observations. First, for all three M-LOCKs, the average accuracy on certified images is comparable to the average accuracy of the baseline CNN network, ensuring the performance of legitimate users . Second, all three M-LOCKs misclassified more than 90\% of uncertified signs, protecting the model from despicable attackers embezzle.

Table \ref{tab:main_result}  reports results for the protection effect of M-LOCK using the specific pattern certificate motif. 
As with the using multi pixels certificate, the precision of the M-LOCK on certified images is only marginally lower than precision of the baseline CNN on clean images. And more importantly, the correct accuracy of M-LOCK on clean(uncertified) images is only 1.3\%, meaning that the M-LOCK active protect success rate higher than 98\% when attackers utilize the flaws of model maintainer in system security protection to illegally obtain access rights.


\section{5. Conclusion}
In this paper, we propose a builtin, universal deep neural network protection sceme Model-Lock (M-LOCK) with several training strategies to prevent malicious thieves from stealing models and obtaining available performance actively when you are away.M-LOCK does not depend on any special or designed structure of models, making it able to be expanded to all existing models and different tasks with insignificant performance affecting.It can even be combined with hardware device features ,which makes software and hardware highly integrated as one thing to employ DLaaS more flexibly and conveniently.M-LOCK is evaluated for various DNN architectures and datasets.We also demonstrated the feasibility and effectiveness of the proposed scheme.




\bibliography{model_lock}

\begin{thebibliography}{13}
\providecommand{\natexlab}[1]{#1}
\providecommand{\url}[1]{\texttt{#1}}
\providecommand{\urlprefix}{URL }
\expandafter\ifx\csname urlstyle\endcsname\relax
  \providecommand{\doi}[1]{doi:\discretionary{}{}{}#1}\else
  \providecommand{\doi}{doi:\discretionary{}{}{}\begingroup
  \urlstyle{rm}\Url}\fi

\bibitem[{Adi and Carsten~Baum(2018)}]{Adi2018}
Adi, Y.; and Carsten~Baum, Moustapha~Cisse, B. P. J.~K. 2018.
\newblock Turning Your Weakness Into a Strength: Watermarking Deep Neural
  Networks by Backdooring.
\newblock In \emph{The Advanced Computing Systems Association (USENIX)}.

\bibitem[{Aniruddha~Saha(2020)}]{Aniruddha2020}
Aniruddha~Saha, Akshayvarun~Subramanya, H.~P. 2020.
\newblock Hidden Trigger Backdoor Attacks.
\newblock In \emph{Association for the Advancement of Artificial Intelligence
  Conference on Artificial Intelligence}.

\bibitem[{Erwan Le~Merrer(2019)}]{Merrer2019}
Erwan Le~Merrer, Patrick~Pérez, G.~T. 2019.
\newblock Adversarial frontier stitching for remote neural network
  watermarking.
\newblock \emph{Neural Computing and Applications} .

\bibitem[{Guowen~Xu(2020)}]{Guowen2020}
Guowen~Xu, Hongwei~Li, Y. Z. X. L. R. H. D. X.~S. 2020.
\newblock A Deep Learning Framework Supporting Model Ownership Protection and
  Traitor Tracing.
\newblock \emph{International Conference on Parallel and Distributed Systems
  (ICPADS)} .

\bibitem[{Jialong~Zhang(2018)}]{Jialong2018}
Jialong~Zhang, Zhongshu~Gu, J. J. H. W. M. P. S. H. H. I. M.~M. 2018.
\newblock Protecting Intellectual Property of Deep Neural Networks with
  Watermarking.
\newblock In \emph{ACM Symposium on Information, Computer and Communications
  Security}.

\bibitem[{Krizhevsky(2009)}]{CIFAR}
Krizhevsky, A. 2009.
\newblock Learning multiple layers of features from tiny images.
\newblock Technical report.

\bibitem[{Manaar~Alam(2020)}]{Manaar2020}
Manaar~Alam, Sayandeep~Saha, D. M. S.~K. 2020.
\newblock Deep-Lock: Secure Authorization for Deep Neural Networks.

\bibitem[{Matthew~Tancik(2020)}]{Matthew2020}
Matthew~Tancik, Ben~Mildenhall, R.~N. 2020.
\newblock StegaStamp: Invisible Hyperlinks in Physical Photographs.
\newblock In \emph{Proceedings of the IEEE/CVF Conference on Computer Vision
  and Pattern Recognition (CVPR)}.

\bibitem[{Potkonjak(2018)}]{Jia2018}
Potkonjak, J. G.~M. 2018.
\newblock Watermarking Deep Neural Networks for Embedded Systems.
\newblock \emph{IEEE International Conference on Computer-Aided Design} .

\bibitem[{Sebastian~Szyller(2020)}]{Szyller2020}
Sebastian~Szyller, Buse Gul~Atli, S. M. N.~A. 2020.
\newblock DAWN: Dynamic Adversarial Watermarking of Neural Networks.

\bibitem[{Xiaoyi~Chen(2020)}]{Xiaoyi2020}
Xiaoyi~Chen, Ahmed~Salem, M. B. S. M. Y.~Z. 2020.
\newblock BadNL: Backdoor Attacks Against NLP Models.

\bibitem[{Yusuke~Uchida(2017)}]{Uchida2017}
Yusuke~Uchida, Yuki~Nagai, S. S. S.~S. 2017.
\newblock Embedding Watermarks into Deep Neural Networks.
\newblock In \emph{ACM International Conference on Multimedia Retrieval
  (ICMR)}.

\bibitem[{Zhicong~Yan(2021)}]{Zhicong2021}
Zhicong~Yan, Gaolei~Li, Y. T. J. W. S. L. M. C. H. V.~P. 2021.
\newblock DeHiB: Deep Hidden Backdoor Attack on Deep Semi-supervised Learning
  via Adversarial Pertubation.
\newblock In \emph{Association for the Advancement of Artificial Intelligence
  Conference on Artificial Intelligence}.

\end{thebibliography}

\end{document}